\documentclass[aps,prd,twocolumn,showkeys,amsmath,amssymb]{revtex4}
\usepackage{graphicx}
\usepackage{epstopdf}   
\usepackage{multirow}
\usepackage{subfigure}
\usepackage{extarrows}
\usepackage{feynmf}
\usepackage{enumitem}
\usepackage[colorlinks,citecolor=blue,anchorcolor=red,menucolor=red,linkcolor=red,filecolor=red,runcolor=red,urlcolor=blue,frenchlinks=red]{hyperref}
\usepackage{color}

\newcommand{\feynp}[1]{#1\kern-0.45em/}

\def\FF(s){\left[(\alpha+\beta)m_c^2-\alpha\beta s\right]}
\def\HH(s){\left[m_c^2-\alpha(1-\alpha) s\right]}
\def\KK(s){\left[\gamma m_c^2-\gamma(1-\gamma) s\right]}

\allowdisplaybreaks[3]

\begin{document}

\title{Fierz analyses on the decay properties of two- and three-gluon glueballs}

\author{Wei-Han Tan}
\author{Hua-Xing Chen}
\email{hxchen@seu.edu.cn}
\affiliation{
School of Physics, Southeast University, Nanjing 211189, China
}

\begin{abstract}
The Fierz rearrangement, based on the various internal symmetries of hadrons, can be used to study their decay properties in a largely model-independent way. In this Letter we apply this method to calculate the relative branching ratios of two-gluon glueballs with \(J^{PC} = 0^{++}/0^{-+}\) and three-gluon glueballs with \(J^{PC} = 0^{++}/1^{+-}\). In total, we derive nearly one hundred ratios for these glueballs. Our results suggest that the \(f_0(1710)\) and \(\eta(2370)\) likely contain a significant gluon component, whereas the gluon component in \(f_0(1500)\) appears to be small. Furthermore, we propose observing the three-gluon glueball with \(J^{PC} = 0^{++}\) in the \(\pi\pi\omega\) and \(K\bar{K}\phi\) channels, and the three-gluon glueball with \(J^{PC} = 1^{+-}\) in the \(\pi\pi\omega\), \(\pi\pi\phi\), and \(K\bar{K}\phi\) channels. This study enhances our understanding of the gluonic structure of exotic hadrons and will assist future experimental searches in high-energy physics.
\end{abstract}

\maketitle

$\\$
{\it Introduction} --- Glueballs, being color-singlet states formed solely through the self-interaction of gluon fields, are of significant importance for the study of non-perturbative Quantum Chromodynamics (QCD)~\cite{pdg}. According to relevant theoretical predictions, the mass of the lightest scalar glueball is approximately \(1.6~\mathrm{GeV}\), while that of the lightest pseudoscalar glueball is around \(2.3~\mathrm{GeV}\)~\cite{Carlson:1982er,DeGrand:1975cf,Isgur:1983wj,Isgur:1984bm,Szczepaniak:1995cw,Llanes-Estrada:2005bii, Mathieu:2008bf,Chen:2005mg,Forkel:2007ru,Li:2013oda,Brunner:2015yha,Michael:1988jr,Yamanaka:2019yek,Richards:2010ck,Ye:2012gu,Morningstar:1999rf,Meyer:2004gx,Gregory:2012hu,Athenodorou:2020ani,Sarantsev:2021ein,Karch:2006pv,Chen:2021bck}. Consequently, several glueball candidates have been discovered experimentally over the past several decades, including the \(f_0(1500)\), \(f_0(1710)\), and \(\eta(2370)\), among others. Extensive theoretical studies have been devoted to interpreting the nature of these states, and we refer to the review papers~\cite{Klempt:2007cp,Crede:2008vw,Mathieu:2008me,Meyer:2010ku,Meyer:2015eta,Ochs:2013gi,Brambilla:2014jmp,Sonnenschein:2016pim,Briceno:2017max,Guo:2017jvc,Bass:2018xmz,Ketzer:2019wmd,Roberts:2021nhw,Fang:2021wes,Jin:2021vct,Gross:2022hyw,Chen:2022asf,Luo:2025sns} and references therein for detailed discussions. However, most of these studies are model-dependent, making it difficult to definitively identify these states as glueballs.

In this Letter we apply the Fierz rearrangement~\cite{Fierz:1937wjm} to study the decay properties of pure glueball states. This method, grounded in the internal symmetries of hadrons, is largely model-independent, as it relies on the fundamental symmetries of QCD. A key property of glueballs, if they exist, is that they couple equally to mesons composed of quarks with different flavors. Furthermore, if glueballs mix with meson states composed of quarks, the mixing coefficients should be comparable for mesons containing quarks of different flavors~\cite{Amsler:1995tu,Klempt:2021wpg,Qin:2017qes}. The model-independent nature of the Fierz rearrangement offers a distinct advantage in describing this property. By relying on the universal symmetries of QCD, it provides a robust framework for understanding glueball interactions without the need for model-dependent assumptions, ensuring its applicability across a wide range of scenarios.

\begin{figure*}[]
\begin{center}
\subfigure[]{
\scalebox{0.4}{\includegraphics{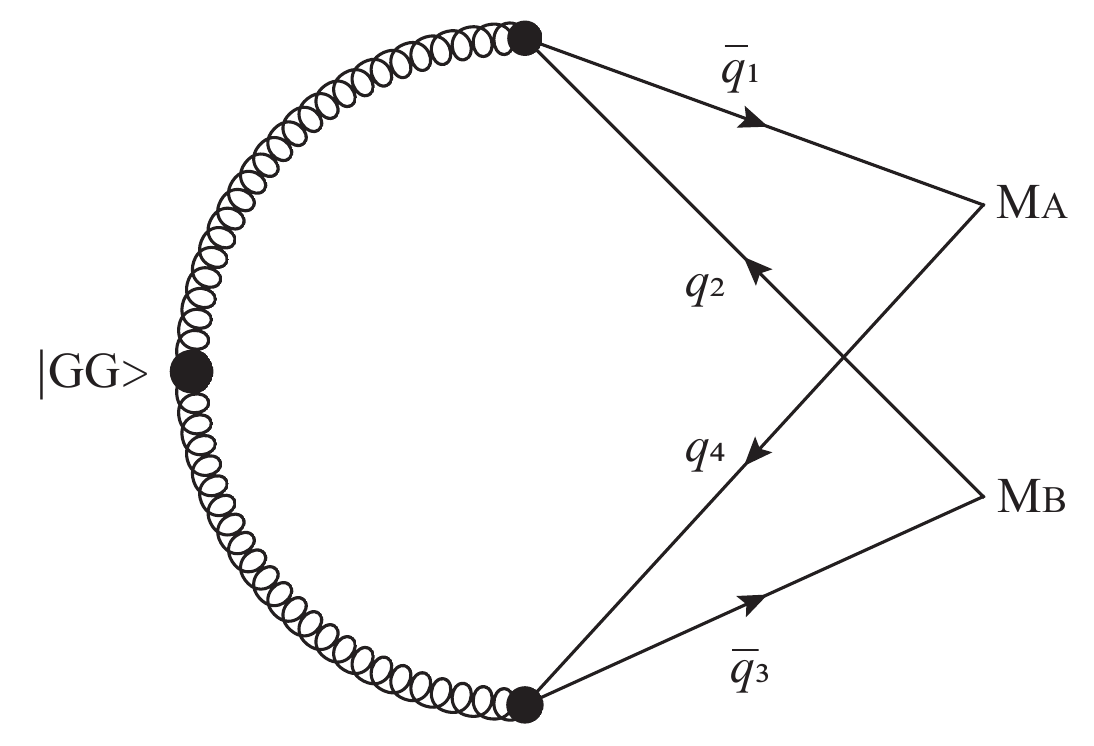}}}~~~~~
\subfigure[]{
\scalebox{0.4}{\includegraphics{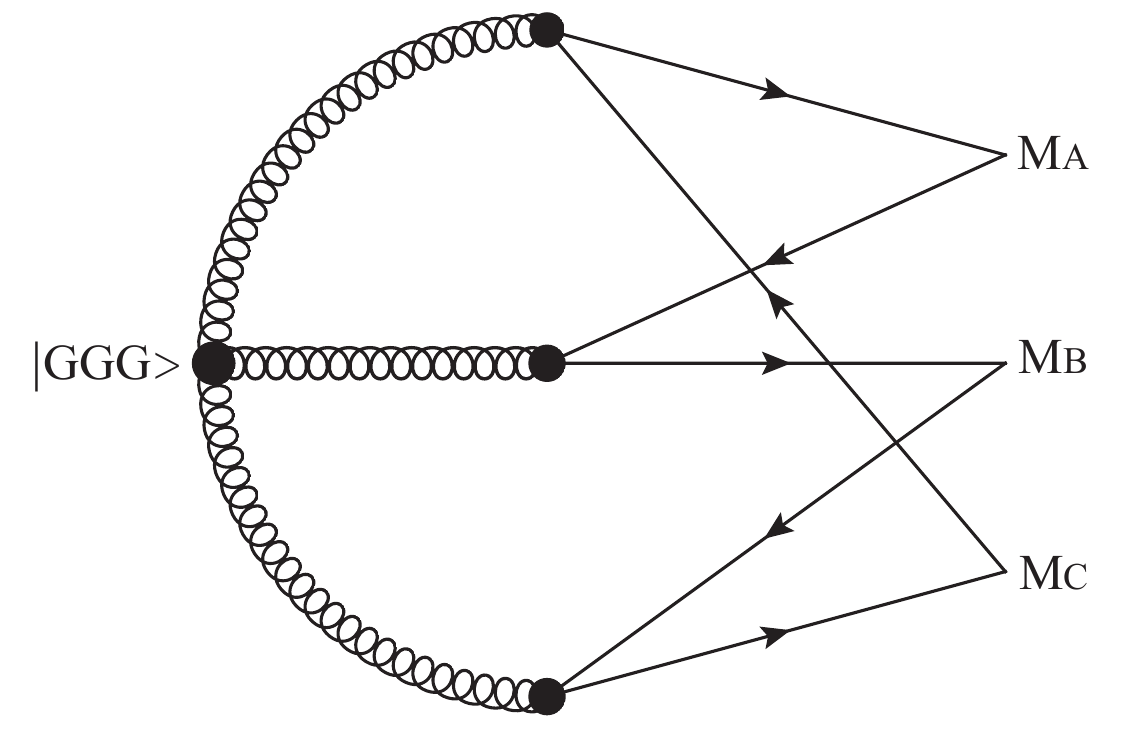}}}~~~~~
\end{center}
\caption{Diagram (a) illustrates the schematic diagram of a two-gluon glueball decaying into two mesons via a single Fierz rearrangement, while Diagram (b) depicts the schematic diagram of a three-gluon glueball decaying into three mesons through two successive Fierz rearrangements.}
\label{fig:decay}
\end{figure*}

We first apply the Fierz rearrangement to investigate the two-gluon glueballs with \(J^{PC} = 0^{++}\) and \(0^{-+}\). As depicted in Fig.~\ref{fig:decay}(a), a two-gluon glueball decays via the excitation of its two gluon fields into two vector quark--antiquark currents. We describe this process through a single Fierz rearrangement by interchanging the quark fields \(q_2\) and \(q_4\). We examine eight decay channels of the two-gluon glueball with \(J^{PC} = 0^{++}\) and subsequently derive seven relative branching ratios. Our results suggest that the \(f_0(1710)\) is a promising candidate, whereas the \(f_0(1500)\) is not favored, based on the relative branching ratios of its decays into the \(K\bar{K}\), \(\pi\pi\), and \(\eta\eta\) channels~\cite{pdg,BESIII:2015rug,BESIII:2018ubj}. We also derive twenty relative branching ratios for the two-gluon glueball with \(J^{PC} = 0^{-+}\), and the results indicate that the \(\eta(2370)\) is a strong candidate. This structure was observed in the \(f_0(980)\eta^\prime\) channel~\cite{BESIII:2023wfi}, and we propose further investigation in the \(\phi\phi\), \(\omega\omega\), and \(\phi\omega\) channels.

\begin{figure*}[]
\begin{center}
\scalebox{1.05}
{\includegraphics{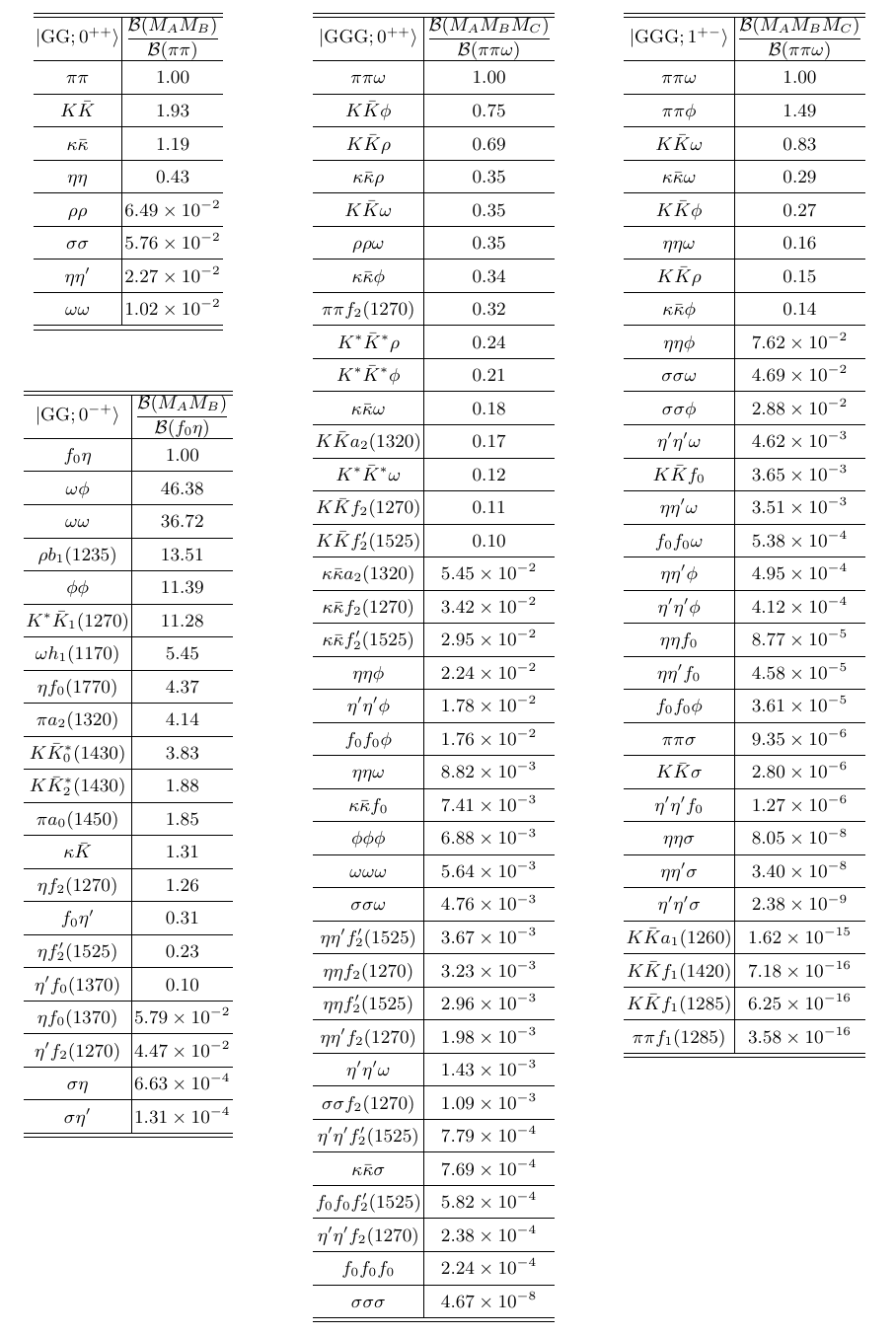}}
\end{center}
\caption{Relative branching ratios of two-gluon glueballs with \(J^{PC}=0^{++}/0^{-+}\) and three-gluon glueballs with \(J^{PC}=0^{++}/1^{+-}\), derived via the Fierz rearrangement, where \(\sigma\) denotes \(f_0(500)\), \(\kappa\) denotes \(K_0^*(700)\), \(f_0\) denotes \(f_0(980)\), and \(K^*\) denotes \(K^*(892)\).}
\label{fig:result}
\end{figure*}

If two-gluon glueballs exist, three-gluon glueballs might also exist, and there have been some experimental signals~\cite{D0:2020tig}. As depicted in Fig.~\ref{fig:decay}(b), we can describe their decay processes through two successive Fierz rearrangements. We apply this method to investigate three-gluon glueballs with \(J^{PC} = 0^{++}\) and \(1^{+-}\), for which we derive thirty-seven and twenty-nine relative branching ratios, respectively. We propose observing the \(J^{PC} = 0^{++}\) state in the \(\pi\pi\omega\) and \(K\bar{K}\phi\) channels, and the \(J^{PC} = 1^{+-}\) state in the \(\pi\pi\omega\), \(\pi\pi\phi\), and \(K\bar{K}\phi\) channels. The relevant relative branching ratios are derived and summarized in Fig.~\ref{fig:result}. This study provides a theoretical foundation for designing targeted search strategies in both current and future high-luminosity experiments, such as BESIII, GlueX, LHCb, and PANDA, among others.

$\\$
{\it Glueball interpolating currents} --- The interpolating currents for two- and three-gluon glueballs have been systematically constructed in Ref.~\cite{Chen:2021bck}. In the present study we investigate two-gluon glueballs with $J^{PC}=0^{++}/0^{-+}$ and three-gluon glueballs with $J^{PC}=0^{++}/1^{+-}$, denoted by $|\mathrm{GG};\,0^{++}/0^{-+}\rangle$ and $|\mathrm{GGG};\,0^{++}/1^{+-}\rangle$, respectively. The corresponding interpolating currents are given by
\begin{align}
J_0 &= g_s^2 G^{\mu \nu}_i G_{\mu \nu}^i  \, , 
\\
\widetilde{J}_0 &= g_s^2 G^{\mu \nu}_i \widetilde{G}_{\mu \nu}^i  \, , 
\\
\eta_0 &= f^{ijk}g_s^3 G^{\mu \nu}_i G_{j,\nu\rho}  G_{k,\mu}^{\rho}\, , 
\\
\eta_1^{\alpha\beta} &= d^{ijk}g_s^3 G^{\mu \nu}_i G_{j,\mu\nu}  G_{k}^{\alpha\beta}\, ,
\end{align}
where $i,j,k$ are color indices (running from 1 to 8), and $\mu,\nu,\rho,\alpha,\beta$ are Lorentz indices; $f^{ijk}$ and $d^{ijk}$ are the antisymmetric and symmetric structure constants, respectively. The gluon field-strength tensor $G_{\mu\nu}^{i}$ can be expressed in terms of the gauge field $A_{\mu}^{i}$ as
\begin{equation}
G_{\mu\nu}^{i} = \partial_{\mu} A_{\nu}^{i} - \partial_{\nu} A_{\mu}^{i} + g_sf^{ijk} A_{\mu}^{j}A_{\nu}^{k} \, .
\end{equation}
So does its dual field $\widetilde{G}_{\mu\nu}^i =G^{i,\rho\sigma} \times \epsilon_{\mu\nu\rho\sigma}/2$.

We assume that a glueball decays via the excitation of its gluon fields into vector quark--antiquark currents,
\begin{equation}
A_{\mu}^{i} \rightarrow \lambda^{i}_{ab}  \times \bar{q}^{a} \gamma_{\mu} q^b\, ,
\end{equation}
where $\lambda^{i}_{ab}$ are the Gell--Mann matrices, with $a$ and $b$ as color indices (running from 1 to 3). Accordingly, the decay processes of two- and three-gluon glueballs are depicted in Fig.~\ref{fig:decay}, which will be discussed separately as follows.

$\\$
{\it Fierz rearrangement on two-gluon currents} --- As an example, we study the decay properties of the two-gluon glueball $|\mathrm{GG};\,0^{++}\rangle$ through the current $J_0$. We need to perform the Fierz rearrangement once, but in both color and Lorentz spaces, as schematically illustrated below:
\begin{align}
|\mathrm{GG};0^{++}\rangle &\xleftrightarrow{~~~~~} J_0 = G^{\mu \nu}_i \times G_{\mu \nu}^i \times g_s^2
\label{eq:expand} \\
\nonumber &\xrightarrow{~~~~~} ({M_{A}})_{\alpha}^{i} \times ({M_{B}})_{\beta}^{i} \times \mathbb{C}_1 + \cdots \\
\nonumber &\xrightarrow{~~~~~} \lambda^{i}_{ab} \lambda^{i}_{cd} \times \bar{q}_1^{a} \gamma_{\alpha} q_2^b \times \bar{q}_3^{c} \gamma_{\beta} q_4^d \times \mathbb{C}_2 + \cdots \\
\nonumber &\xrightarrow{\rm color} \delta_{ad} \delta_{cb} \times \bar{q}_1^{a} \gamma_{\alpha} q_2^b \times \bar{q}_3^{c} \gamma_{\beta} q_4^d \times \mathbb{C}_3 + \cdots \\
\nonumber &\xrightarrow{\rm Fierz} \delta_{ad} \delta_{cb} \times \bar{q}_1^{a} \Gamma_{1} q_4^d \times \bar{q}_3^{c} \Gamma_{2} q_2^b \times \mathbb{C}_4 + \cdots \\
\nonumber &\xrightarrow{~~~~~} [\bar{q}_1^{a} \Gamma_{1} q_4^a]_A \times [\bar{q}_3^b \Gamma_{2} q_2^b]_B \times \mathbb{C}_4 + \cdots \, .
\end{align}
Here, $\mathbb{C}_{1,2,3,4}$ and $\Gamma_{1,2}$ are the relevant coefficients and Dirac structures to be determined. The quark fields $\bar{q}_1^{a}$ and $q_2^{b}$ carry the same flavor, and likewise, $\bar{q}_3^{c}$ and $q_4^{d}$ carry the same flavor. The subscripts $1,2,3,4$, labeling the four quark fields, and the subscripts $M_A,M_B$, labeling the two meson components, are introduced for convenience in the discussion below. In particular, we employ the color identity
\begin{equation}
\lambda^{i}_{ab}\lambda^{i}_{cd} = 2\delta_{ad}\delta_{cb}-\frac{2}{3}\delta_{ab}\delta_{cd}\, ,
\end{equation}
and we also require the Fierz rearrangement in Lorentz space, for which the relevant formulae can be found in Ref.~\cite{Chen:2016qju}.

The detailed result of Eq.~(\ref{eq:expand}) will be presented in our forthcoming paper. In total, there are forty-eight terms, and we display only seven of them:
\begin{eqnarray}
J_0 &=& +\frac{1}{4}[\bar{u}_1^a \gamma_{\mu}\gamma_5 d_4^a]_A [\bar{d}_3^b \gamma_{\mu}\gamma_5 u_2^b]_B \times(q_A^{\nu}+q_B^{\nu})^2 
\\ \nonumber & & +\frac{1}{2}[\bar{u}_1^a \gamma_{\mu}\gamma_5 d_4^a]_A [\bar{d}_3^b \gamma_{\nu}\gamma_5 u_2^b]_B \times (q_A^{\mu}+q_B^{\mu})(q_A^{\nu}+q_B^{\nu})
\\ \nonumber & & +\frac{3}{4}[\bar{u}_1^a \gamma_5 d_4^a]_A [\bar{d}_3^b \gamma_5 u_2^b]_B\times (q_A^2+q_B^2+2q_A \cdot  q_B)
\\ \nonumber & & +\frac{1}{4}[\bar{u}_1^a \gamma_{\mu} d_4^a]_A [\bar{d}_3^b \gamma_{\mu} u_2^b]_B \times(q_A^{\nu}+q_B^{\nu})^2
\\ \nonumber & & +\frac{1}{2}[\bar{u}_1^a \gamma_{\mu} d_4^a]_A [\bar{d}_3^b \gamma_{\nu} u_2^b]_B \times (q_A^{\mu}+q_B^{\mu})(q_A^{\nu}+q_B^{\nu})
\\ \nonumber & & -\frac{1}{2}[\bar{u}_1^a \sigma_{\mu\alpha} d_4^a]_A [\bar{d}_3^b \sigma_{\nu\alpha} u_2^b]_B \times (q_A^{\mu}+q_B^{\mu})(q_a^{\nu}+q_b^{\nu})
\\ \nonumber & & +\frac{1}{8}[\bar{u}_1^a \sigma_{\mu\nu} d_4^a]_A [\bar{d}_3^b \sigma_{\mu\nu} u_2^b]_B \times (q_A^2+q_B^2+2q_A \cdot  q_B)\, .
\label{eq:fierz}
\end{eqnarray}
Here, we have replaced the partial derivative operator acting on a quark-antiquark current with the momentum of the corresponding meson component; \textit{e.g.}, $\partial^{\mu}[\bar{u}_1^a \gamma_5 d_4^a]_A = p_A^{\mu} \times [\bar{u}_1^a \gamma_5 d_4^a]_A$. This result will be used to evaluate the relative branching ratio $\mathcal{B}(|\mathrm{GG}; 0^{++}\rangle \to \rho^+ \rho^- ) / \mathcal{B}(|\mathrm{GG}; 0^{++}\rangle \to \pi^+ \pi^- )$ as described below.

$\\$
{\it Decay analysis on two-gluon glueballs} --- The couplings of various quark--antiquark currents to meson states are well established in the literature~\cite{pdg,Tan:2025nir}; \textit{e.g.},
\begin{align}
\langle 0 | \bar u_a i \gamma_5 d_a | \pi^-(q) \rangle &= \lambda_{\pi} \, ,
\label{eq:decaycp2}
\\ \nonumber \langle 0 | \bar u_a \gamma_\mu \gamma_5 d_a | \pi^-(q) \rangle &= i q_\mu f_{\pi} \, ,
\\ \nonumber \langle 0 | \bar u_a \gamma_\mu  d_a | \rho^-(q, \epsilon) \rangle &= m_{\rho} f_{\rho} \epsilon_{\mu} \, ,
\\ \nonumber \langle 0 | \bar u_a \sigma_{\mu\nu} d_a | \rho^-(q, \epsilon) \rangle &= i  f_{\rho}^T(q_{\mu}\epsilon_{\nu}-q_{\nu}\epsilon_{\mu}) \, .
\end{align}
Using these couplings together with Eq.~(\ref{eq:fierz}), we obtain:
\begin{enumerate}

\item The decay amplitude of $|\mathrm{GG};\,0^{++}\rangle$ into the $\pi^+\pi^-$ channel receives contributions from both
$[\bar{u}_{a}\gamma_5 d_a]\,[\bar{d}_{b}\gamma_5 u_b]$ and
$[\bar{u}_{a}\gamma_{\mu_1}\gamma_5 d_a]\,[\bar{d}_{b}\gamma_{\mu_2}\gamma_5 u_b]$.
Altogether, it can be written as
\begin{eqnarray}
\mathcal{M}_{\pi\pi}  &\propto
& - { 3 \over 4}  \lambda_{\pi}^2  (q_1^2 + q_2^2 + 2q_1 \cdot
 q_2)
\\ \nonumber & & - { 3 \over 4} f_{\pi}^2  (q_1^2+q_2^2) (q_1 \cdot q_2) 
\\ \nonumber & & - { 1 \over 2} f_{\pi}^2 (q_1^2 \times q_2^2 + 2(q_1 \cdot q_2)^2)\, ,
\end{eqnarray}
where $q_{1,2}$ are the momenta of the two pions.

\item The decay amplitude of $|\mathrm{GG};\,0^{++}\rangle$ into the $\rho^+\rho^-$ channel receives contributions from both
$[\bar{u}_{a}\gamma_{\mu_1} d_a]\,[\bar{d}_{b}\gamma_{\mu_2} u_b]$ and
$[\bar{u}_{a}\sigma_{\mu_1\nu_1} d_a]\,[\bar{d}_{b}\sigma_{\mu_2\nu_2} u_b]$.
Altogether, it can be written as
\begin{eqnarray}
\mathcal{M}_{\rho\rho} &\propto& + { 1 \over 2} m_{\rho}^2 f_{\rho}^2  (\epsilon(q_1)\cdot q_2)  (\epsilon(q_2)\cdot q_1)
 \\ \nonumber & & +{ 1 \over 4} m_{\rho}^2 f_{\rho}^2  (\epsilon(q_1)\cdot \epsilon(q_2))  (q_1^2+q_2^2+2q_1\cdot q_2)
\\ \nonumber & & + { 1 \over 4} (f_{\rho}^T)^2  (\epsilon(q_1)\cdot \epsilon(q_2)) (q_1^2+q_2^2)(q_1\cdot q_2)
\\ \nonumber & & + { 1 \over 2} (f_{\rho}^T)^2  (\epsilon(q_1)\cdot \epsilon(q_2)) q_1^2 \times q_2^2 
\\ \nonumber & & - { 1 \over 4} (f_{\rho}^T)^2 (\epsilon(q_1)\cdot q_2)  (\epsilon(q_2)\cdot q_1) (q_1^2+q_2^2)\, ,
\end{eqnarray}
where $q_{1,2}$ are the momenta of the two $\rho$ mesons, and $\epsilon_{1,2}$ are their polarization vectors.

\end{enumerate}
The amplitudes above can be used to compute the partial decay widths of $|\mathrm{GG};\,0^{++}\rangle$ into the $\pi^+\pi^-$ and $\rho^+\rho^-$ channels. After factoring out the overall common factor appearing in all amplitude expressions, we obtain the relative branching ratio between these two channels:
\begin{equation}
\frac{\mathcal{B}(|\mathrm{GG};\,0^{++}\rangle \to \rho \rho )}{\mathcal{B}(|\mathrm{GG};\,0^{++}\rangle \to \pi \pi )} = 6.49\times10^{-2} \, .
\end{equation}
Here, the input masses and decay constants of the two-gluon glueball, as well as those of the \( \rho \) and \( \pi \) mesons, will introduce uncertainties into the aforementioned results. Since we primarily adopt the experimentally measured values in this study, the relative errors in the branching ratios across different decay channels are relatively small for two-gluon glueballs, while they are expected to be somewhat larger for three-gluon glueballs.

Similarly, we compute the relative branching ratios for various decay channels of two-gluon glueballs with $J^{PC} = 0^{++}$ and $0^{-+}$:
\begin{itemize}

\item Our results suggest that the \(f_0(1710)\) is a good candidate for a two-gluon glueball with \(J^{PC} = 0^{++}\). Using its mass, 1733~\text{MeV}~\cite{pdg}, as input, we derive relative branching ratios for eight decay channels, which are summarized in Fig.~\ref{fig:result}. In particular, the \(K\bar{K}\) and \(\pi\pi\) channels exhibit relatively large branching fractions. The relative branching ratio of these two channels is found to be
\begin{eqnarray}
\frac{\mathcal{B}(\mathrm{GG};0^{++} \to K \bar{K})}{\mathcal{B}(\mathrm{GG};0^{++} \to \pi\pi)} &=& 1.93 \, ,
\end{eqnarray}
which is consistent with the experimental measurements listed in Ref.~\cite{pdg}. Regarding the \(f_0(1500)\), its relative branching ratios measured in experiments—especially those for the \(K\bar{K}\), \(\pi\pi\), and \(\eta\eta\) channels—differ significantly from the results obtained in this work, suggesting that the \(f_0(1500)\) is unlikely to be a pure glueball and may, in fact, contain only a minor gluon component.

\item Our results suggest that the \(\eta(2370)\) is a good candidate for a two-gluon glueball with $J^{PC}=0^{-+}$. Using its mass, 2377~\text{MeV}~\cite{pdg}, as input, we derive relative branching ratios for twenty decay channels, which are also summarized in Fig.~\ref{fig:result}. In particular, the \(\phi\phi\), \(\omega\omega\), and \(\phi\omega\) channel exhibit relatively large branching fractions. Since the \(\eta(2370)\) has been observed in the \(f_0(980)\eta^\prime\) channel~\cite{BESIII:2023wfi}, it is interesting to further investigate it in these three decay channels.

\end{itemize}

$\\$
{\it Decay analysis on three-gluon glueballs} --- The decay properties of three-gluon glueballs can be studied in a similar manner. However, two successive Fierz rearrangements are required, which makes the calculation considerably more involved. In particular, we need the following color identities:
\begin{eqnarray}
f^{ijk}\lambda_{i}^{ab}\lambda_{j}^{cd}\lambda_{k}^{ef} & =&
+ 2i\delta^{af}\delta^{bc}\delta^{de}-2i\delta^{ad}\delta^{be}\delta^{cf}\, ,
\\ \nonumber d^{ijk}\lambda_{i}^{ab}\lambda_{j}^{cd}\lambda_{k}^{ef} &=& +2\delta^{af}\delta^{bc}\delta^{de}+2\delta^{ad}\delta^{be}\delta^{cf}
\\  \nonumber && -\frac{4}{3}\delta^{ab}\delta^{de}\delta^{cf}
 -\frac{4}{3}\delta^{af}\delta^{cd}\delta^{be}
\\  && -\frac{4}{3}\delta^{ad}\delta^{bc}\delta^{ef}+\frac{8}{9}\delta^{ab}\delta^{cd}\delta^{ef}\, .
\end{eqnarray}

We apply the Fierz rearrangement to compute the relative branching ratios for various decay channels of three-gluon glueballs with $J^{PC}=0^{++}$ and $1^{+-}$:
\begin{itemize}

\item The mass of the three-gluon glueball with $J^{PC}=0^{++}$ was estimated in Ref.~\cite{Chen:2021bck} to be 4210~\text{MeV}. Using this value as input, we derive relative branching ratios for thirty-seven decay channels, which are summarized in Fig.~\ref{fig:result}. In particular, the \(\pi\pi\omega\) and \(K\bar{K}\phi\) channels exhibit relatively large branching fractions.

\item The mass of the three-gluon glueball with $J^{PC}=1^{+-}$ was estimated in Ref.~\cite{Chen:2021bck} to be 3190~\text{MeV}. Using this value as input, we derive relative branching ratios for twenty-nine decay channels, which are summarized in Fig.~\ref{fig:result}. In particular, the \(\pi\pi\omega\), \(\pi\pi\phi\), and \(K\bar{K}\phi\) channels exhibit relatively large branching fractions.

\end{itemize}

$\\$
{\it Summary} --- In this Letter we employ the Fierz rearrangement to compute the relative branching ratios for various decay channels of two-gluon glueballs with \(J^{PC} = 0^{++}\) and \(0^{-+}\). Our results suggest that the \(f_0(1710)\) and \(\eta(2370)\) likely contain a significant gluon component, whereas the gluon contribution in \(f_0(1500)\) appears to be relatively small. We extend this approach to three-gluon glueballs with \(J^{PC} = 0^{++}\) and \(1^{+-}\). For the \(J^{PC} = 0^{++}\) state, we propose observing it in the \(\pi\pi\omega\) and \(K\bar{K}\phi\) channels, while the \(J^{PC} = 1^{+-}\) state can be studied in the \(\pi\pi\omega\), \(\pi\pi\phi\), and \(K\bar{K}\phi\) channels.

Our findings provide a theoretical framework to assist the search for glueball states in high-luminosity experiments, such as BESIII, GlueX, LHCb, and PANDA. These results also contribute to the ongoing efforts to identify exotic hadrons and enhance our understanding of QCD dynamics.

\section*{Acknowledgments}

We thank Wen-Ying Liu and Ding-Kun Lian for useful discussions.
This project is supported by
the National Natural Science Foundation of China under Grant No.~12075019,
the Jiangsu Provincial Double-Innovation Program under Grant No.~JSSCRC2021488,
the SEU Innovation Capability Enhancement Plan for Doctoral Students No.~CXJHSEU25139,
and
the Fundamental Research Funds for the Central Universities.

\end{document}